\documentclass{optica-article}
\usepackage{lineno}
\usepackage{amsmath}
\usepackage{diagbox}
\usepackage{xcolor}

\journal{opticajournal}
\articletype{Research Article}
\DeclareMathOperator{\sech}{sech}
\begin{document}
	
	\title{Stable dissipative topological solitons in media with even-order
		dispersion}
	\author{Xianan Sun,\authormark{1,$\dag$} Dongdong Wang,\authormark{1,2,$%
			\dag$} Dumitru Mihalache,\authormark{3} David Laroze,\authormark{4} Boris A.
		Malomed, \authormark{5,6} and Pengfei Li \authormark{1,2,*}}
	\address{\authormark{1}Department of Physics, Taiyuan Normal University, Jinzhong, Shanxi 030619, China\\
		\authormark{2}Shanxi Key Laboratory for Intelligent Optimization Computing and Blockchain Technology, Taiyuan Normal University, Jinzhong, Shanxi 030619, China\\
		\authormark{3}Horia Hulubei National Institute of Physics and Nuclear Engineering, Magurele, Bucharest RO-077125, Romania\\
		\authormark{4}Instituto de Alta Investigaci\'{o}n, Universidad de Tarapac\'{a}, Casilla 7D, Arica, Chile\\
		\authormark{5}Department of Physical Electronics, School of Electrical Engineering, Faculty of Engineering, and Center for Light-Matter Interaction, Tel Aviv University, Tel Aviv 69978, Israel\\
		\authormark{6}School of Physics and Optoelectronic Engineering, Foshan University, Foshan 528000, China\\		
		\authormark{\dag}The authors contributed equally to this work.\\
		\authormark{*}lpf281888@gmail.com}
	
	\begin{abstract}
		The formation and dynamics of dissipative solitons are governed by the interplay of dispersion and nonlinearity, alongside the gain-loss balance. While it is known that dispersion orders beyond the conventional second order support stable solitons, the possibility of the creation of dissipative topological solitons in this setting was not explored. Here, we reveal a class of stable dissipative solitons enabled by the pure-high-even-order dispersion (PHEOD), with dispersion orders $m=4$, $6$, $8$, and $10$. The resulting dissipative topological solitons (DTSs) and moving topological solitons (MTSs) are dark-type structures featuring a zero-amplitude point that separates two CW backgrounds with opposite signs. In particular, the PHEOD solitons feature oscillatory tails, which are not present in the conventional dark states. The linear-stability analysis confirms full robustness of the PHEOD-supported DTSs and MTSs. We also explore collision dynamics of MTSs, whose repulsive interactions are inelastic, leading either to rebound, accompanied by emission of dispersive radiation, or complete fusion. These findings establish PHEOD as a means for the generation of new species of dissipative solitons, and broaden the variety of nonlinear modes available to experimental studies in nonlinear optics.
	\end{abstract}
	
	
	\section{Introduction}
	
	\label{Sec I}
	
	Dissipative solitons are localized structures that emerge in
	driven nonconservative systems far from equilibrium. Unlike conservative
	solitons, whose formation relies on the balance between nonlinear effects
	and dispersion or diffraction, dissipative solitons rely upon the additional
	balance between gain and loss, which is necessary to compensate energy
	dissipation and sustain the localized modes under the action of the drive
	\cite{DS-Book1,DS-Book2}. The inclusion of the gain-loss balance makes
	dissipative solitons fixed-point (\textit{attractor}) solutions of the
	corresponding complex Ginzburg-Landau equation (CGLE), in contrast to the
	standard nonlinear Schr\"{o}dinger equation (NLSE) for conservative solitons
	\cite{DS-Book3}, which exists in continuous families, rather than isolated
	attractors. In nonlinear optics, CGLEs model a wide range of systems,
	including passively mode-locked lasers \cite{NatPhoton-6-84} and optical
	parametric oscillators \cite%
	{ol-192050,ol-20-695,PRE-53-5520,PRA-77-023814,NatPhoton-20-310}. Among
	these models, two generic types are directly forced CGLEs, alias
	Lugiato-Lefever equations \cite{PRL-Book3-58-2209}, and
	parametrically-driven damped NLSEs, in which parametric forcing compensates
	linear losses \cite%
	{BBK,Pelinovsky,PRL-83-2568,PRL-90-054103,PRE-79-016605,NatPhoton-15-857,NatPhoton-18-617,PRE-114-014223}.
	
	A broader class of localized nonlinear waves emerges when the
	dispersion is represented by a single pure-high-even-order dispersion
	(PHEOD) term, rather than the conventional second-order one. In this case,
	solitons arise from the balance between the nonlinear self-focusing and
	PHEOD, with pure-quartic solitons (PQSs) corresponding to the lowest-order
	member of the family. Dispersion engineering has provided experimental
	realization of PQSs in photonic-crystal waveguides and mode-locked laser
	systems employing intracavity spectral pulse shaping \cite%
	{NC-7-10427,LPR-14-492,LPR-20-e01851}. More recently, experiments have
	extended this concept to sixth-, eighth-, and tenth-order dispersion,
	producing the respective solitons \cite{PRR-3-013166}. PHEOD solitons have
	also been shown to form stable bound states, revealing increasingly complex
	localized-wave dynamics beyond the PQS regime \cite%
	{LSA-13-101,CSF-204-117742}. An important feature of this hierarchy is its
	specific energy scaling: for dispersion order $m$, the soliton energy $E$
	varies with the pulse duration $\tau _{0}$ as $E\propto \tau _{0}^{-(m-1)}$.
	This characteristic enables the PHEOD solitons to break the energy
	constraint inherent to conventional second-order-dispersion solitons.
	
	The study of dissipative higher-order solitons has extended from
	PQSs to a broader class of PHEOD setups. PQSs exhibit relatively flat
	spectra and, as mentioned above, the favorable energy scaling, making them
	attractive for high-power ultrashort-pulse generation \cite%
	{OE-30-22066,OE-32-47882,OE-33-10129}. Dissipative PQSs have subsequently
	been demonstrated in microring resonators within the Lugiato--Lefever
	framework \cite{OL-44-3086,OL-47-2438}, and their robustness against the Raman
	self-frequency shift has been explored \cite{OL-46-993,OL-50-2848,OL-50-3357}%
	. Beyond the quartic dispersion, dissipative PHEOD solitons have been
	realized through the loss compensation provided by the ac or parametric
	drive \cite{OL-50-4262,OL-51-3217}. These advances establish PHEOD as a
	viable platform for the creation of dissipative solitons.
	
	In this work, we investigate dissipative topological solitons supported by PHEOD. These states feature localized phase defects connecting two asymptotic continuous-wave (CW) backgrounds with a phase difference of $\pi$ \cite{RMP-51-591}. The term ``topological soliton'' refers to this kink-like structure, following the terminology adopted in optics \cite{Book-Topological-Solitons,OL-22-970,JOB-1-133}. The linear-stability analysis and direct numerical simulations show that the topological PHEOD solitons remain \emph{completely stable} throughout the parameter regions examined, indicating their strong intrinsic robustness. In addition to representing a new type of stable dissipative solitons, these modes may be potential candidates for optical telecommunications and data-processing schemes \cite{NatPhoton-11-102}, as well as high-Q sources of soliton frequency combs \cite{Nature-652-76}.
	
	\section{The model and methods}
	
	\label{Sec II}
	
	We consider one-dimensional dissipative solitons
	in parametrically driven dispersive cavities with the Kerr nonlinearity \cite%
	{NatPhoton-15-857,NatPhoton-18-617}. The evolution of the intracavity field envelope $A (t_s,T)$ obeys NLSE containing the PHEOD term (we address the cases of $m=4,6,8,10$) \cite{PRR-3-013166,LSA-13-101}. We establish a necessary theoretical model incorporating the PHEOD, Kerr nonlinearity, parametric driving and loss. The equation is written as
	
	\begin{equation}
		i t_R \frac{\partial A}{\partial t_s}+i^m\frac{\beta_m}{m!}\frac{\partial^m A}{\partial T^m}+\kappa |A|^2 A+\Delta A =\rho A^*-i\Gamma A, \label{1a}
	\end{equation}
	where the fast time $T$ denotes the retarded time in the reference frame moving at the group velocity, $t_s$ is the slow-time variable,
	and $t_R$ represents the cavity round-trip time. Furthermore,
	$\beta_m$ is the $m$th-order dispersion coefficient, $\kappa$
	is the nonlinearity coefficient, $\Delta$ denotes the cavity
	detuning, $\rho$ represents the strength of the parametric drive,
	and $\Gamma$ is the effective loss coefficient.
	
	To obtain a dimensionless form of the governing equation, we introduce the following scaled variables and dimensionless parameters:
	\begin{equation}
		t=\frac{|\Delta|}{t_R}t_s,\qquad
		T_0=\left(\frac{|\beta_m|}{m!|\Delta|}\right)^{1/m},
		\qquad
		\tau=\frac{T}{T_0}-Vt, \label{1b}
	\end{equation}
	and
	\begin{equation}
		\psi=\sqrt{\frac{|\kappa|}{|\Delta|}}A,\qquad
		h=\frac{\rho}{|\Delta|},\qquad
		\gamma=\frac{\Gamma}{|\Delta|}. \label{1c}
	\end{equation}
	With these definitions, the governing equation
	can be recast into the normalized parametrically driven
	NLSE
	\begin{equation}
		i\frac{\partial \psi }{\partial t}-iV\frac{\partial \psi }{\partial \tau }%
		-\left( -1\right) ^{m/2}\frac{\partial ^{m}\psi }{\partial \tau ^{m}}-|\psi
		|^{2}\psi +\psi =h\psi ^{\ast }-i\gamma \psi,  \label{1}
	\end{equation}
	where the scaled moving coordinate is used to accommodate moving solitons with velocity $V$. Parameters $\gamma \geq 0$ and $h>0$ denote, respectively, the normalized cavity loss and parametric-drive amplitude, with $\ast $ standing for the complex conjugate. It is well known that parametrically driven systems with the second-order dispersion support bright dissipative solitons, a part of which may be stable \cite{NatPhoton-15-857,NatPhoton-18-617}.
	
	It is worth noting that, under the sign convention adopted here, all the considered cases correspond to negative high-even-order dispersion coefficients, $\beta_m<0$, and the self-defocusing Kerr nonlinearity in Eq. (\ref{1a}). In Eq. (\ref{1}), the alternating signs arise from the differential representation of the even-order dispersion, and do not indicate a change between anomalous and normal dispersion. Thus, all the cases share the same relative dispersion–nonlinearity sign supporting dark modes.
	
	To obtain stationary profiles in the comoving frame, we seek solutions of Eq. (\ref{1}) in the form of
	\begin{equation}
		\psi (\tau )=\psi _{r}(\tau )+i\psi _{i}(\tau ),  \label{DTS}
	\end{equation}%
	where $\psi _{r}$ and $\psi _{i}$ denote the real and imaginary components
	of the stationary field. Substitution of Eq. (\ref{DTS}) into Eq. (\ref{1})
	yields the coupled equations:
	\begin{equation}
		V\frac{d\psi _{i}}{d\tau }-\left( -1\right) ^{m/2}\frac{d^{m}\psi _{r}}{%
			d\tau ^{m}}-|\psi _{r}+i\psi _{i}|^{2}\psi _{r}+\psi _{r}-h\psi _{r}-\gamma
		\psi _{i}=0,  \label{Stationary1}
	\end{equation}%
	\begin{equation}
		-V\frac{d\psi _{r}}{d\tau }-\left( -1\right) ^{m/2}\frac{d^{m}\psi _{i}}{%
			d\tau ^{m}}-|\psi _{r}+i\psi _{i}|^{2}\psi _{i}+\psi _{i}+h\psi _{i}+\gamma
		\psi _{r}=0.  \label{Stationary2}
	\end{equation}%
	The stationary soliton profiles are obtained numerically by solving Eqs. (%
	\ref{Stationary1}) and (\ref{Stationary2}) using a
	Newton--conjugate-gradient method \cite{JCP-228-7007,YJK-Book}. This
	approach allows both quiescent and moving localized states to be computed
	within the same stationary formulation, with $V=0$ corresponding to the
	quiescent limit.
	
	To assess the linear stability of the resulting solutions, we introduce a
	small perturbation according to
	\begin{equation}
		\psi _{\mathrm{pert}}\left( \tau ,t\right) =\left[ u(\tau )+iv(\tau )\right]
		e^{\delta t},  \label{Perturbation}
	\end{equation}
	where $u(\tau)$ and $v(\tau)$ are the real and imaginary components of the
	perturbation eigenmode, respectively, and $\delta=\delta_R+i\delta_I$ is the
	associated eigenvalue. The real part $\delta_R=\mathrm{Re}(\delta)$
	determines the perturbation growth rate. Substitution of Eq. (\ref%
	{Perturbation}) into Eq. (\ref{1}) and linearization about the stationary
	solution yield the eigenvalue problem

	\begin{equation}
		\left(
		\begin{array}{cc}
			L_{11} & L_{12} \\
			L_{21} & L_{22}%
		\end{array}
		\right) \left(
		\begin{array}{c}
			u \\
			v%
		\end{array}
		\right) =\delta \left(
		\begin{array}{c}
			u \\
			v%
		\end{array}
		\right) ,  \label{LSA-Eq}
	\end{equation}
	where
	\begin{eqnarray}
		L_{11} &=&2\psi _{r}\psi _{i}+V\frac{d}{d\tau }-\gamma,  \label{Operator-L11} \\
		L_{12} &=&+\left( -1\right) ^{m/2}\frac{d^{m}}{d\tau ^{m}}+\left( \psi
		_{r}^{2}+3\psi _{i}^{2}\right) -1-h,  \label{Operator-L12} \\
		L_{21} &=&-\left( -1\right) ^{m/2}\frac{d^{m}}{d\tau ^{m}}-\left( 3\psi
		_{r}^{2}+\psi _{i}^{2}\right) +1-h,  \label{Operator-L21} \\
		L_{22} &=&-2\psi _{r}\psi _{i}+V\frac{d}{d\tau }-\gamma.  \label{Operator-L22}
	\end{eqnarray}
	A localized solution is unstable if the linearized operator possesses at
	least one eigenvalue with $\delta_R>0$; otherwise, it is stable within the
	linear approximation. We compute the eigenvalue spectrum using a
	Fourier-collocation discretization \cite{YJK-Book}, which transforms Eq. (%
	\ref{LSA-Eq}) into a finite-dimensional matrix eigenvalue problem for the
	Fourier coefficients of the perturbation modes $u$ and $v$.
	
	\section{Results and discussion}
	
	\label{Sec VI} To characterize the structure of the soliton solutions, we
	first analyze a pair of the underlying CW states, which Eq. (\ref{1})
	produces as
	\begin{equation}
		\psi _{\pm }=A_{\pm }e^{i\Theta _{\pm }},  \label{2}
	\end{equation}%
	with
	\begin{equation}
		A_{\pm }^{2}=1\mp \sqrt{h^{2}-\gamma ^{2}},\Theta _{\pm }=\frac{1}{2}\arccos
		\left( \pm \sqrt{1-\frac{\gamma ^{2}}{h^{2}}}\right) ,  \label{3}
	\end{equation}%
	under the condition of $h^{2}\geq \gamma ^{2}$. To test the CW
	states against plane-wave perturbations, we introduce perturbation
	amplitudes $\left( u,v\right) $ with real wavenumber $k$ and complex growth
	rate $\delta $ as
	\begin{equation}
		\psi _{\mathrm{CW}}\left( \tau ,t\right) =\psi _{\pm }+\left( ue^{\delta
			t+ik\tau }+ive^{\delta t-ik\tau }\right) .  \label{Perturbation1}
	\end{equation}%
	The linearization about $\psi _{\pm }$ with respect to $\left(
	u,v\right) $ yields the dispersion relation
	\begin{equation}
		\delta =-\gamma +iVk\pm i\sqrt{Z},  \label{dispersion-relation}
	\end{equation}%
	\begin{equation}
		Z\equiv \left( 1-2A_{\pm }^{2}-k^{m}\right) ^{2}+A_{\pm }^{2}\left( A_{\pm
		}^{2}-2\right) -h^{2},  \label{Z}
	\end{equation}%
	the CW state being stable provided that condition $\mathrm{Re}(\delta )\leq 0
	$ holds for all real $k$. Because Eq. (\ref{3}) yields $A_{+}^{2}<1$, the CW
	state $\psi _{+}$ is always unstable, while $\psi _{-}$, which has $%
	A_{-}^{2}>1$, is stable.
	
	A related analysis predicts the shape of tails of stationary solutions
	existing on top of the CW states, such as the dissipative topological
	solitons (DTSs) considered below. In this case, the ansatz is taken as $\psi
	=\psi _{\pm }+\psi _{1}\left( \tau \right) $, where $\psi _{1}(\tau )$ is
	the small tail component (cf. the stationary version of Eq. (\ref%
	{Perturbation1}), with $\delta =0$). Splitting $\psi _{-}$ and $\psi
	_{1}(\tau )$ into real and imaginary parts, substituting this in Eqs. (\ref%
	{Stationary1}) and (\ref{Stationary2}), linearizing the equations with
	respect to $\psi _{1}(\tau )$, and finally looking for the solutions as $%
	\left( \psi _{1}\right) _{r}$, $\left( \psi _{1}\right) _{i}\sim \exp \left(
	\lambda \tau \right) $ leads to the equation that determines the tail's
	damping rate $\lambda $:
	\begin{equation}
		\lambda ^{2m}+2(-1)^{m/2}\left( 2A_{\pm }^{2}-1\right) \lambda
		^{m}+V^{2}\lambda ^{2}-2\gamma V\lambda +3A_{\pm }^{4}-4A_{\pm
		}^{2}-2hA_{\pm }^{2}\cos (2\Theta _{\pm })+1+\gamma ^{2}-h^{2}=0,
		\label{lambda}
	\end{equation}
	where $A_{\pm }$ and $\Theta _{\pm }$ are defined as per Eq. (\ref{3}).
	Although functions $\left( \psi _{1}\right) _{r}\ $and $\left( \psi
	_{1}\right) _{i}$ are defined as real variables, complex solutions of Eq. ( %
	\ref{lambda}) for $\lambda $ make sense, predicting the possibility of
	oscillatory decaying tails (cf. Ref. \cite{OL-47-2438}), as shown below in
	Fig. \ref{fig1}.
	
	\subsection{Stable dissipative topological solitons (DTSs)}
	
	\label{Sec VIa}
	
	\begin{figure}[h]
		\centering\vspace{0cm} \includegraphics[width=1\linewidth]{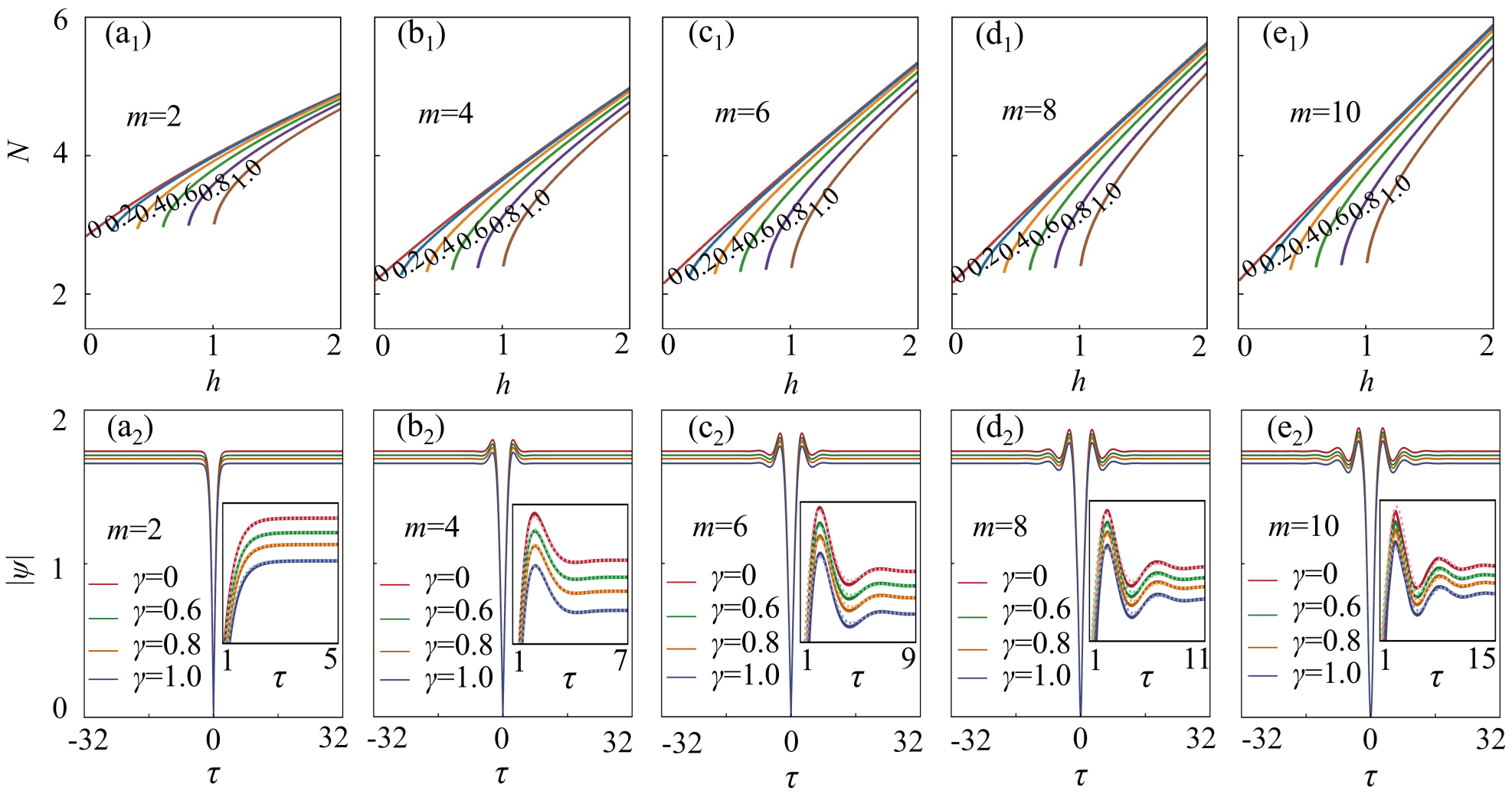} \vspace{-0.5cm}
		\caption{The energy (\protect\ref{N}) of stationary DTS
			solutions, for varying values of the loss coefficient $\protect\gamma $
			(attached as labels to the plotted branches) and PHEOD order $m$, versus the parametric-drive's strength $h$. Panels $(\text{a}_{2})$-$(\text{e}_{2})$
			exhibit characteristic profiles of $\left\vert \protect\psi (\protect\tau )\right\vert $, produced, at fixed $h=2$, for different values of $\protect\gamma $. The insets present the comparison between the oscillatory decaying tails, as predicted by Eq. (\protect\ref{lambda}) and produced by the numerical DTS solutions (dotted and solid lines, respectively).}
		\label{fig1}
	\end{figure}
	
	We begin by considering the quiescent DTSs described by Eqs. (\ref%
	{Stationary1}) and (\ref{Stationary2}) in the stationary limit, $V=0$. In
	the conventional quadratic-dispersion case ($m=2$), this solution recovers
	the familiar dark-soliton state \cite{PRL-90-054103}:
	\begin{equation}
		\psi (\tau )_{\mathit{m=2}}=iA\tanh \left( \frac{A}{\sqrt{2}}\tau \right)
		e^{i\Theta },  \label{DTS1-k2}
	\end{equation}
	\begin{equation}
		A^{2}=1+\sqrt{h^{2}-\gamma ^{2}},\qquad \Theta =\frac{1}{2}\arcsin \left(
		\frac{\gamma }{h}\right) ,  \label{DTS1-k2-A-Theta}
	\end{equation}
	which exist under the above-mentioned condition, $h^{2}>\gamma ^{2}$.
	
	Our findings for DTS families are summarized in Figs. \ref{fig1}$(\text{a}
	_{1})$-\ref{fig1}$(\text{e}_{1})$, by means of dependencies of the
	complementary power (similar to the total power of bright solitons),
	\begin{equation}
		N=\int_{-\tau _{0}}^{+\tau _{0}}\left( |\psi _{\text{CW}}|^{2}-|\psi
		|^{2}\right) d\tau ,~\mathrm{with~~}\tau _{0}=32,  \label{N}
	\end{equation}
	on the drive's strength $h$, for several fixed values of the the loss coefficients values $\gamma $ (starting from $\gamma =0$), and dispersion orders $m=2,4,6,8,10$. The linear stability analysis confirms that all DTS solutions are stable.
	
	Typical examples of their temporal profiles, found at fixed $h=2$ and various values of $\gamma $, are displayed in Figs. \ref{fig1}$(\text{a}
	_{2})$-\ref{fig1}$(\text{e}_{2})$, clearly exhibiting oscillatory decaying
	tails, which is consistent with previous results \cite{PRR-3-013166}. As
	mentioned above, the oscillatory decay corresponds to complex roots of Eq. (\ref{lambda}), $\lambda \equiv \lambda _{r}+i\lambda _{i}$. In the case of $V=0$, Eq. (\ref{lambda}) amounts to the factorized form:
	\begin{equation}
		\left[ \lambda ^{m}-2(-1)^{\frac{m}{2}+1}\left( 1+\sqrt{h^{2}-\gamma ^{2}}
		\right) \right] \left[ \lambda ^{m}-2(-1)^{\frac{m}{2}+1}\left( \sqrt{
			h^{2}-\gamma ^{2}}\right) \right] =0.  \label{lambda_m}
	\end{equation}
	The oscillatory decaying tails of the complex solutions of Eqs. (\ref%
	{Stationary1}) and (\ref{Stationary2}) are determined by eigenvalues $%
	\lambda $ in the form of
	\begin{equation}
		\psi =\psi _{-}+k_{1}e^{\lambda _{r}\tau }\cos (\lambda _{i}\tau +\phi
		_{1})+ik_{2}e^{\lambda _{r}\tau }\cos (\lambda _{i}\tau +\phi _{2}),
		\label{psi_tail}
	\end{equation}
	where the constants $k_{1}$, $k_{2}$, $\phi _{1}$, and $\phi _{2}$ are
	fitting parameters. The magnified view for $1<\tau <15$, displayed in insets
	in Figs. \ref{fig1}$(\text{a}_{2})$-\ref{fig1}$(\text{e}_{2})$ demonstrates full agreement of the analytical prediction (\ref{psi_tail}) with the numerical solution.
	
	Furthermore, recent experiments have demonstrated the feasibility of engineering pure high-even-order dispersion in fiber lasers using the intracavity programmable spectral shaping \cite{LSA-13-101}. Reported net cavity dispersion coefficients include $\beta_4=-2.66\,\mathrm{ps}^{4}$, $\beta_6=-2.128\,\mathrm{ps}^{6}$, $\beta_8=-2.128\,\mathrm{ps}^{8}$, and 	$\beta_{10}=-2.128\,\mathrm{ps}^{10}$. Using these dispersion magnitudes as benchmarks and taking a round-trip phase detuning as $\Delta=0.01\,\mathrm{rad}$, the normalization $T_0=[|\beta_m|/(m!|\Delta|)]^{1/m}$ gives characteristic time scales $\approx$ $1.82$, $0.816$, $0.519$, and $0.377\,\mathrm{ps}$ for $m=4$, $6$, $8$, and $10$, respectively. For the stationary dark solitons with $h=2$ and $\gamma=1$, the corresponding dimensional widths are therefore $\approx$ $3.5$, $1.7$, $1.2$, and $0.87\,\mathrm{ps}$, respectively. These estimates illustrate the physical temporal scales of the predicted dark modes. Note that the reported solitons are domain walls connecting two opposite-phase CW backgrounds in an effectively infinite temporal domain. An isolated wall therefore does not satisfy ordinary periodic boundary conditions for the complex optical field. In a periodic cavity, a reversed-phase domain can instead be bounded by a pair of oppositely oriented walls, allowing the field to recover its initial complex amplitude after one round trip. When sufficiently separated, the individual walls may locally approximate the isolated solutions studied here, although the existence and stability of the complete periodic configuration require special verification. A promising implementation is a phase-stabilized, parametrically driven Kerr fiber cavity incorporating the programmable intracavity dispersion. Reported experiments have demonstrated these driving and dispersion-engineering capabilities, providing a basis for exploring the proposed dark-domain-wall regime.
	
	\begin{figure}[h]
		\centering\vspace{0cm} \includegraphics[width=1\linewidth]{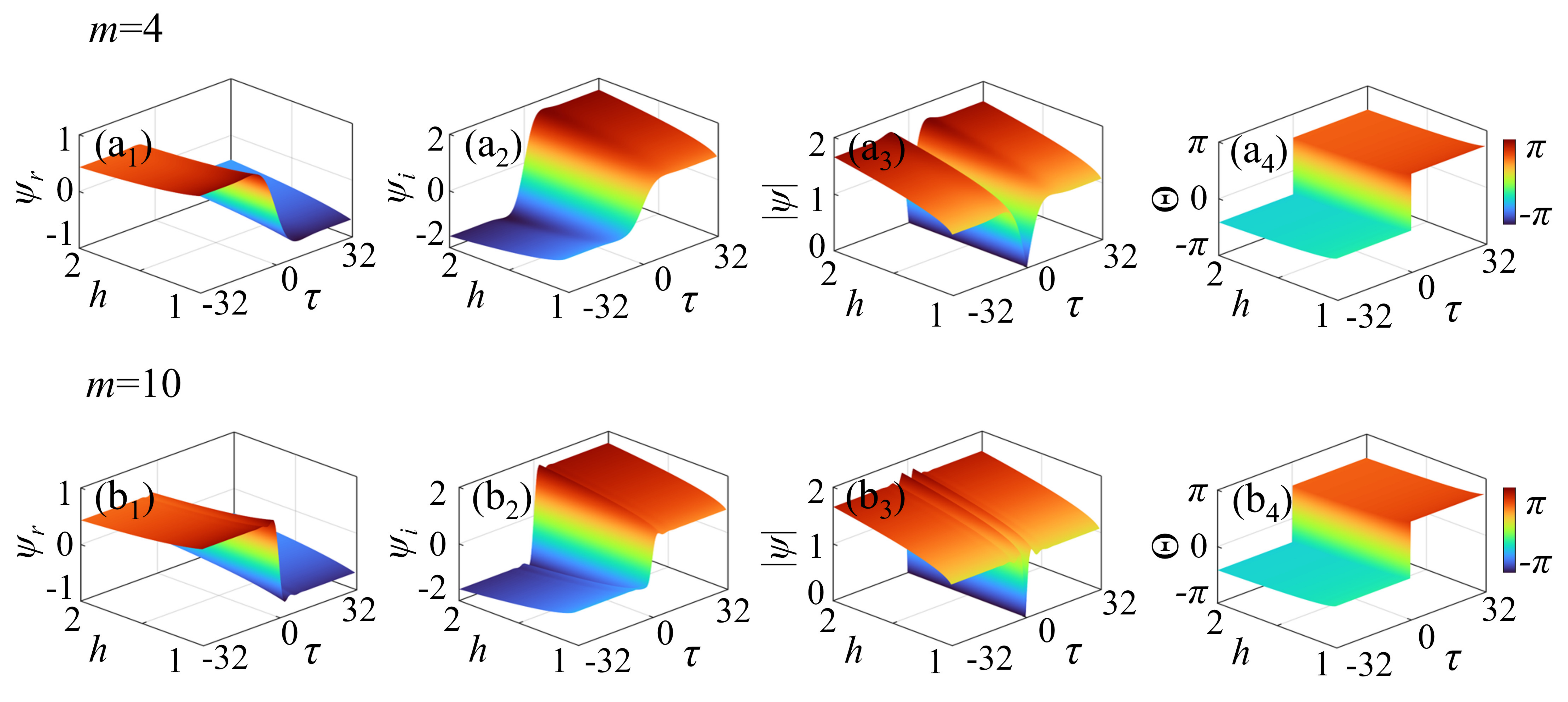} \vspace{%
			-0.5cm}
		\caption{DTS solutions of Eqs. (\protect\ref{Stationary1})
			and (\protect\ref{Stationary2}), with $V=0$, varying with the drive's
			strength $h$ and PHEOD order $m$. Panels $(\text{a}_{1})$-$(\text{b}_{1})$
			and $(\text{a}_{2})$-$(\text{b}_{2})$ trace the real and imaginary field
			components, $\protect\psi_r$ and $\protect\psi_i$, as $h$ changes at fixed $%
			\protect\gamma=1$ for $m=4$ and $m=10$, respectively. Panels $(\text{a}_{3})$%
			-$(\text{b}_{3})$ and $(\text{a}_{4})$-$(\text{b}_{4})$ display,
			respectively, the corresponding field's modulus profiles and phase distributions of the DTSs.}
		\label{fig2}
	\end{figure}
	
	Structural features of quiescent DTSs exhibit a pronounced dependence on the
	parametric-drive's strength $h$. For a fixed loss constant $\gamma =1$,
	Figs. \ref{fig2}$(\text{a}_{1})$-\ref{fig2}$(\text{b}_{1})$ and \ref{fig2}$(
	\text{a}_{2})$-\ref{fig2}$(\text{b}_{2})$ present the variation of the real
	and imaginary field components ($\psi _{r}$ and $\psi _{i}$) as functions of
	$h$ . The corresponding field's modulus profiles are displayed in Figs. \ref%
	{fig2}$(\text{a}_{3})$-\ref{fig2}$(\text{b}_{3})$, which connect to the CW
	background $A_{\mathit{N}}\equiv \pm \sqrt{1+\sqrt{h^{2}-\gamma ^{2}}}$ at $%
	\tau \rightarrow \pm \infty $, provided that the above-mentioned condition, $%
	h^{2}>\gamma ^{2}$, holds. Phase distributions of DTSs are plotted in Figs. %
	\ref{fig2}$(\text{a}_{4})$-\ref{fig2}$(\text{b}_{4})$, revealing that each
	DTS is formed by a pair of half-infinite CW states, with the $\pi $ phase
	shift between them, making it possible to identify the DTS as a topological
	soliton.
	
	\begin{figure}[h]
		\centering\vspace{0cm} \includegraphics[width=1\linewidth]{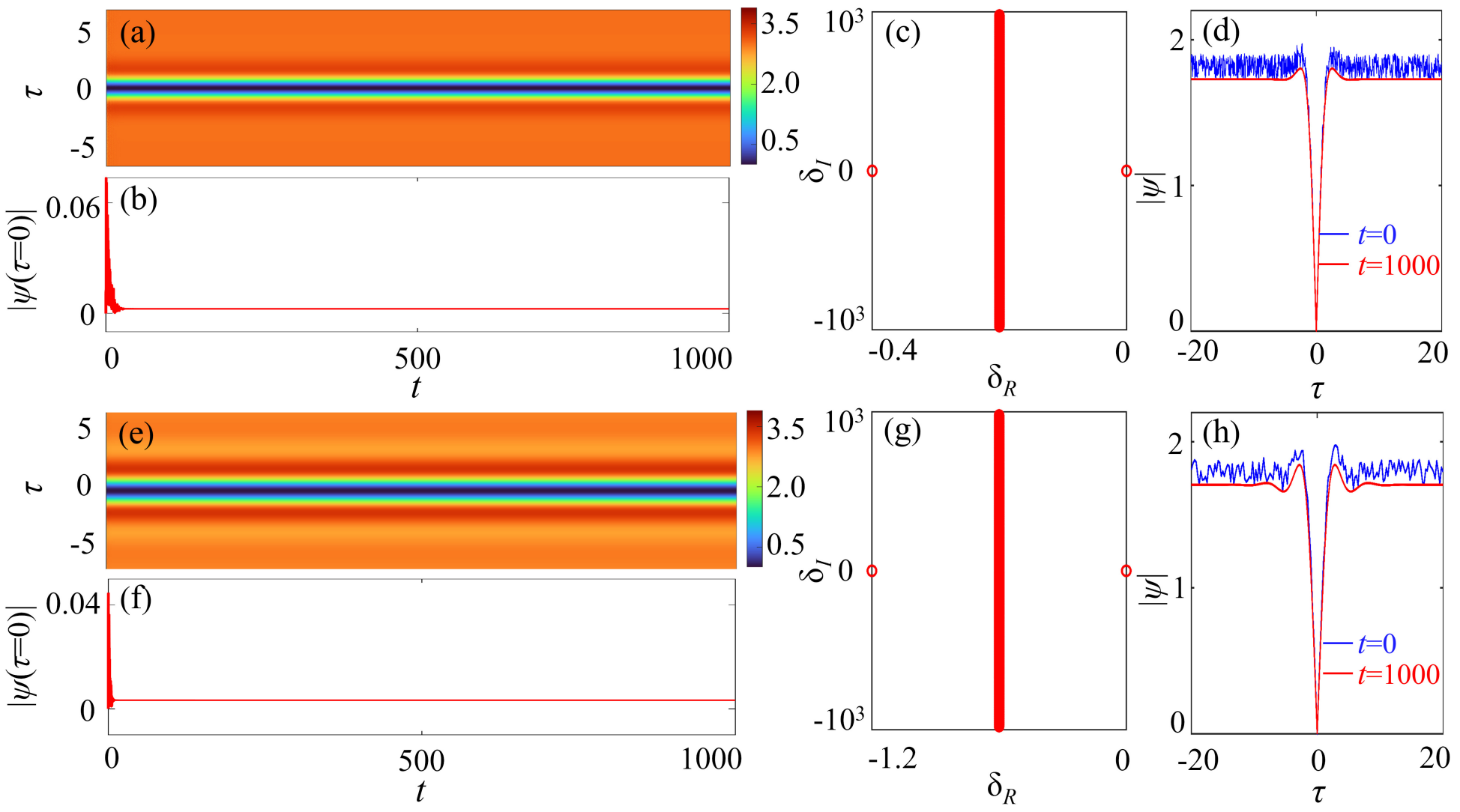} \vspace{%
			-0.5cm}
		\caption{Evolution tests and linear-stability spectra for
			quiescent DTSs. For $m=4$, $h=2$, and $\protect\gamma=0.2$, panel $(\text{a}%
			) $ shows the evolution of a DTS subjected to $10\,\%$ random-amplitude
			noise; panel $(\text{b})$ tracks $|\protect\psi(\protect\tau=0)|$; panel $(%
			\text{c}) $ gives the corresponding stability spectrum; and panel $(\text{d}%
			) $ compares the perturbed input with the profile at $t=1000$. Panels $(%
			\text{e}) $--$(\text{h})$ present the analogous results for $m=8$, $h=2$,
			and $\protect\gamma=0.6$.}
		\label{fig3}
	\end{figure}
	
	Topological solitons are robust states with a localized core, whose temporal profiles are governed by the interplay of the PHEOD and optical nonlinearity. To assess their stability, we performed direct propagation simulations with perturbed initial conditions, $\psi(\tau,0)=\psi_0(\tau)[1+\epsilon r(\tau)]$, where $\psi_0$ is the stationary soliton profile, $\epsilon=0.1$, and $r(\tau)$ are independent real random numbers uniformly distributed over $[0,1]$ at numerical-grid points. This perturbation introduces multiplicative amplitude noise with a relative amplitude of $10\%$, without imposing independent phase noise. Although the random factors are real, the additive field perturbation is, generally, complex because $\psi_0$ is complex, its real and imaginary components being multiplied by the same random factor. The propagation simulations, presented in Fig. \ref{fig3}, provide direct numerical evidence of the robustness of the DTSs against the imposed amplitude perturbations. Panels \ref{fig3}$(\text{a})$ and \ref{fig3}$(\text{e})$ show that the solitons retain their shapes in the course of long evolution. The central amplitude $|\psi (\tau =0)|$, plotted in panels \ref{fig3}$(\text{b})$ and \ref{fig3}$(\text{f})$, quickly settles after the introduction of the perturbation. The spectra in panels \ref{fig3}$(\text{c})$ and \ref{fig3}$(\text{g})$, obtained from Eq. (\ref{LSA-Eq}), are consistent with this behavior. The comparison of the initial and final profiles, displayed in panels \ref{fig3}$(\text{d})$ and \ref{fig3}$(\text{h})$, corroborate that perturbation-induced oscillations gradually decay, without destroying the DTS.

	\subsection{Stable moving topological solitons (MTSs)}
	
	\label{Sec VIb}
	
	\begin{figure}[h]
		\centering\vspace{0cm} \includegraphics[width=1\linewidth]{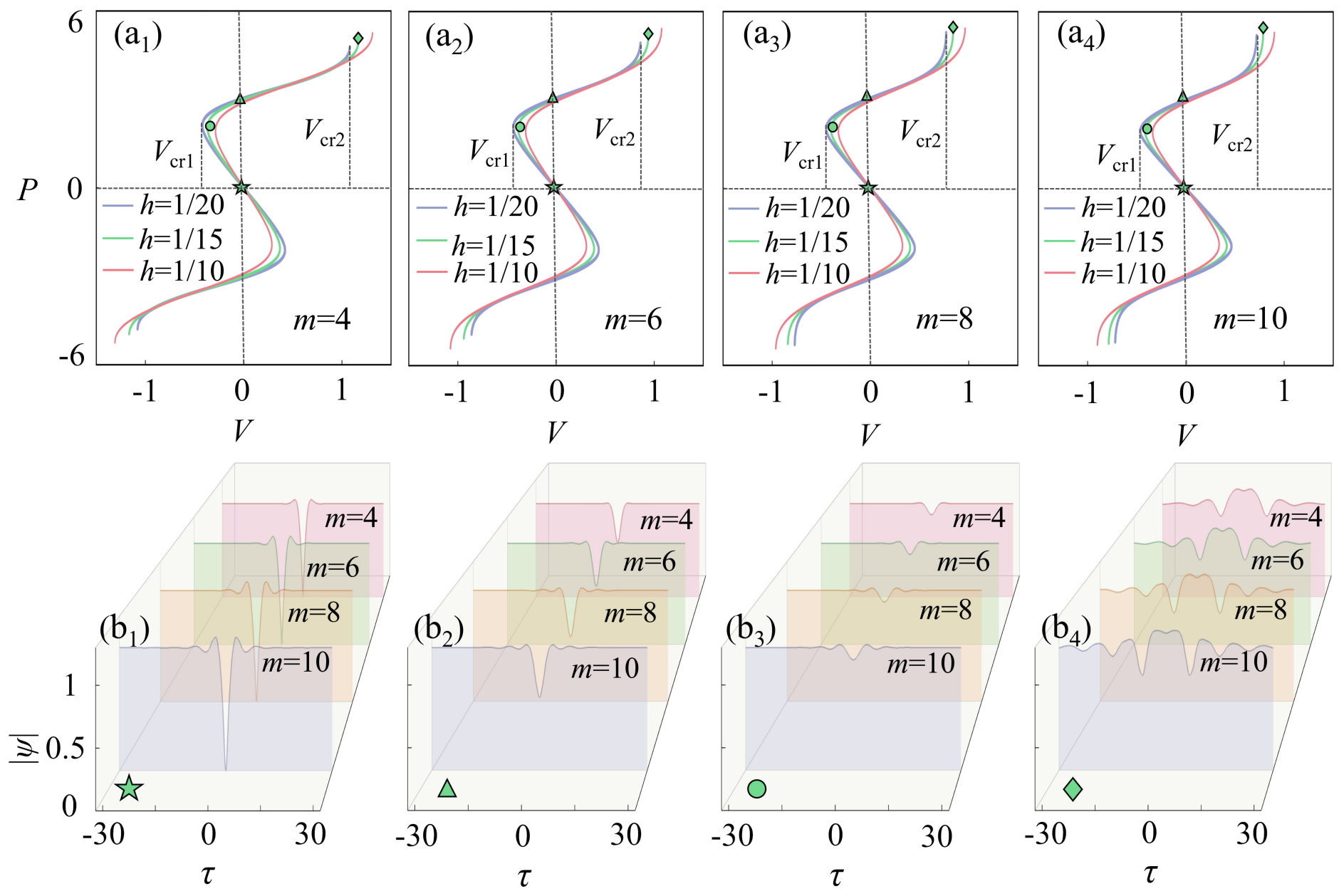} \vspace{%
			-0.5cm}
		\caption{MTS existence curves in the conservative
			limit, $\protect\gamma =0$. Panels $(\text{a}_{1})$--$(\text{a}_{4})$,
			corresponding to $m=4,6,8,$ and $10$, plot the momentum $P$ [Eq. (\protect
			\ref{Momentum})] against velocity $V$ for several values of $h$. Dashed
			vertical lines mark the two critical velocities $V_{\text{cr1}}$ and $V_{%
				\text{cr2}}$ on the upper ($P>0$) branches. Panels $(\text{b}_{1})$--$(\text{%
				b}_{4})$ show representative modulus profiles at $h=1/15$.}
		\label{fig4}
	\end{figure}
	
	The parametric-drive and PHEOD terms break the Galilean
	invariance of the NLSE \cite{PRA-104-043526}, which makes the construction
	of moving solitons a nontrivial problem. To produce moving topological
	solitons solutions of Eq. (\ref{1}), we here consider the lossless case ($%
	\gamma =0$). In the case of the quadratic dispersion, Eqs. (\ref{Stationary1}%
	) and (\ref{Stationary2}) for this conservative system give rise to the
	well-known family of dark-soliton solutions \cite{PRL-90-054103},
	\begin{equation}
		\psi (\tau )_{\mathit{m=2}}=-iA\tanh (\sqrt{2h}\tau )\pm \sqrt{1-3h}\sech(%
		\sqrt{2h}\tau ),  \label{DTS2-k2}
	\end{equation}%
	whose existence is restricted to $0<h<1/3$.
	
	Numerical solutions of Eqs. (\ref{Stationary1}) and (\ref%
	{Stationary2}) yield the PHEOD-supported MTS families summarized in Fig. \ref%
	{fig4}. Panels \ref{fig4}$(\text{a}_{1})$--\ref{fig4}$(\text{a}_{4})$
	represent these families, plotting the momentum
	
	\begin{equation}
		P=\frac{i}{2}\int_{-\infty }^{+\infty }\left( \frac{d\psi ^{\ast }}{d\tau }
		\psi -\frac{d\psi }{d\tau }\psi ^{\ast }\right) d\tau ,  \label{Momentum}
	\end{equation}
	as a function of $V$. In the conservative limit considered here,
	both $P$ and the norm $N$ are conserved.
	\begin{table}[h]
		\caption{The critical velocities $V_{\text{cr1}}$ and $V_{\text{cr2}}$ of
			the upper MTS branches ($P>0$) \ in Figs. (\protect\ref{fig4})(a$_{1}$-a$%
			_{4} $), for different values of the parametric-gain strength $h$ and PHEOD
			orders $m$.}
		\label{table1}\centering
		\setlength{\tabcolsep}{6pt}
		\resizebox{1\linewidth}{!}{
			\begin{tabular}{cccc}
				\hline\hline
				\diagbox[font=\small]{$m$}{$V_\text{cr1}$($P$), $V_\text{cr2}$($P$)}{$h$} & $1/20$ & $1/15$ & $1/10$ \\ \hline
				$4$  &-0.420(2.085), 1.081(4.778)&-0.369(2.023), 1.167(4.924)&-0.287(1.901), 1.312(5.244) \\
				$6$  &-0.436(2.045), 0.862(4.930)&-0.388(1.980), 0.941(5.068)&-0.310(1.877), 1.077(5.415) \\
				$8$  &-0.451(2.034), 0.772(5.292)&-0.404(2.009), 0.842(5.260)&-0.327(1.899), 0.965(5.427) \\
				$10$ &-0.464(1.945), 0.721(5.190)&-0.417(1.932), 0.787(5.243)&-0.339(1.891), 0.902(5.292) \\ \hline\hline
			\end{tabular}
		} \vspace{0.1cm}
	\end{table}
	
	The existence diagrams for MTSs exhibit odd symmetry with respect to the
	velocity $V$, revealing that MTSs exist only within a range $|V|\leq |V_{%
		\mathrm{cr2}}|$, where $V_{\mathrm{cr2}}$ is the critical velocity.
	Furthermore, each upper branch ($P>0$) features another critical velocity, $%
	V_{\mathrm{cr1}}$, which defines the maximum speed of the retrograde motion
	of the MTSs on that branch. Table \ref{table1} reveals the variation of the
	critical velocities. Specifically, $V_{\mathrm{cr2}}$ increases with the
	parametric-gain strength $h$, and decreases as the PHEOD order $m$ rises.
	Further, representative field-modulus profiles of MTSs are presented in
	Figs. \ref{fig4}$(\text{b}_{1})$-\ref{fig4}$(\text{b}_{4})$. Figures \ref%
	{fig4}$(\text{b}_{1})$ and \ref{fig4}$(\text{b}_{2})$ show stationary MTSs
	at zero velocity ($V=0$), while Figs. \ref{fig4}$(\text{ b}_{3})$ and \ref%
	{fig4}$(\text{b}_{4})$ represent MTSs propagating at the critical velocities
	$V_{\mathrm{cr1}}$ and $V_{\mathrm{cr2}}$, respectively. The
	linear-stability analysis further confirms that all these MTS solutions are
	stable.
	
	\begin{figure}[h]
		\centering\vspace{0cm} \includegraphics[width=1\linewidth]{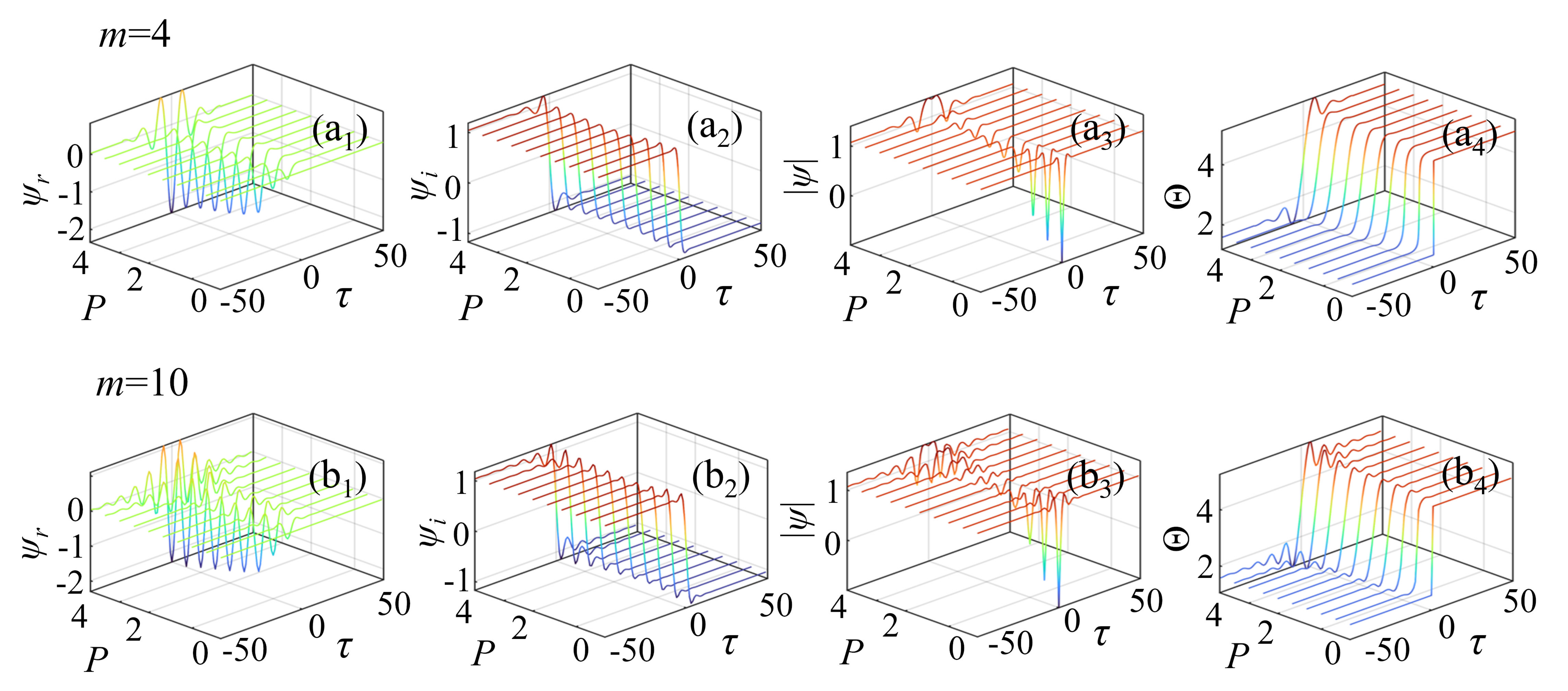} \vspace{%
			-0.5cm}
		\caption{Dependence of MTS profiles on momentum $P
			$ and dispersion order $m$ at fixed $h=1/15$. For $m=4$ and $m=10$, panels $(%
			\text{a}_{1})$, $(\text{b}_{1})$ and $(\text{a}_{2})$, $(\text{b}_{2})$
			display $\protect\psi _{r}$ and $\protect\psi _{i}$, respectively. The
			associated modulus and phase profiles are presented in panels $(\text{a}_{3})
			$--$(\text{b}_{3})$ and $(\text{a}_{4})$--$(\text{b}_{4})$.}
		\label{fig5}
	\end{figure}
	
	Figure \ref{fig5} summarizes how the MTS structure changes with
	momentum $P$ and PHEOD order $m$. At a fixed parametric-gain strength $h$,
	the evolution of the real and imaginary components ($\psi_r$ and $%
	\psi_i$) with $P$ is shown in Figs. \ref{fig5}$(\text{a}_{1})$--\ref{fig5}$(%
	\text{b}_{1})$ and \ref{fig5}$(\text{a}_{2})$--\ref{fig5}$(\text{b}_{2})$.
	The associated field-modulus profiles are displayed in Figs. \ref{fig5}$(%
	\text{ a }_{3})$-\ref{fig5}$(\text{b}_{3})$, which asymptotically merge with
	the CW background ($\pm \sqrt{1+h}$). Distinct from DTSs, the oscillatory
	tails of MTSs are modulated both by PHEOD order $m$ and soliton's momentum $%
	P $, with the modulation amplitude rising with $m$ and $P$. Furthermore, as $%
	P$ approaches its maximum value, corresponding to the critical velocity $V_{
		\mathrm{cr2}}$, the region around the central point $\tau =0$ no longer
	retains the central dip, inherent to conventional dark solitons, with a
	bubble-like structural feature emerging instead. The corresponding phase
	distributions of the MTSs are presented in Figs. \ref{fig5}$(\text{a}_{4})$- %
	\ref{fig5}$(\text{b}_{4})$, confirming that the MTSs arise from two
	half-infinite CW states separated by the $\pi $ phase shift.
	
	\begin{figure}[h]
		\centering\vspace{0cm} \includegraphics[width=1\linewidth]{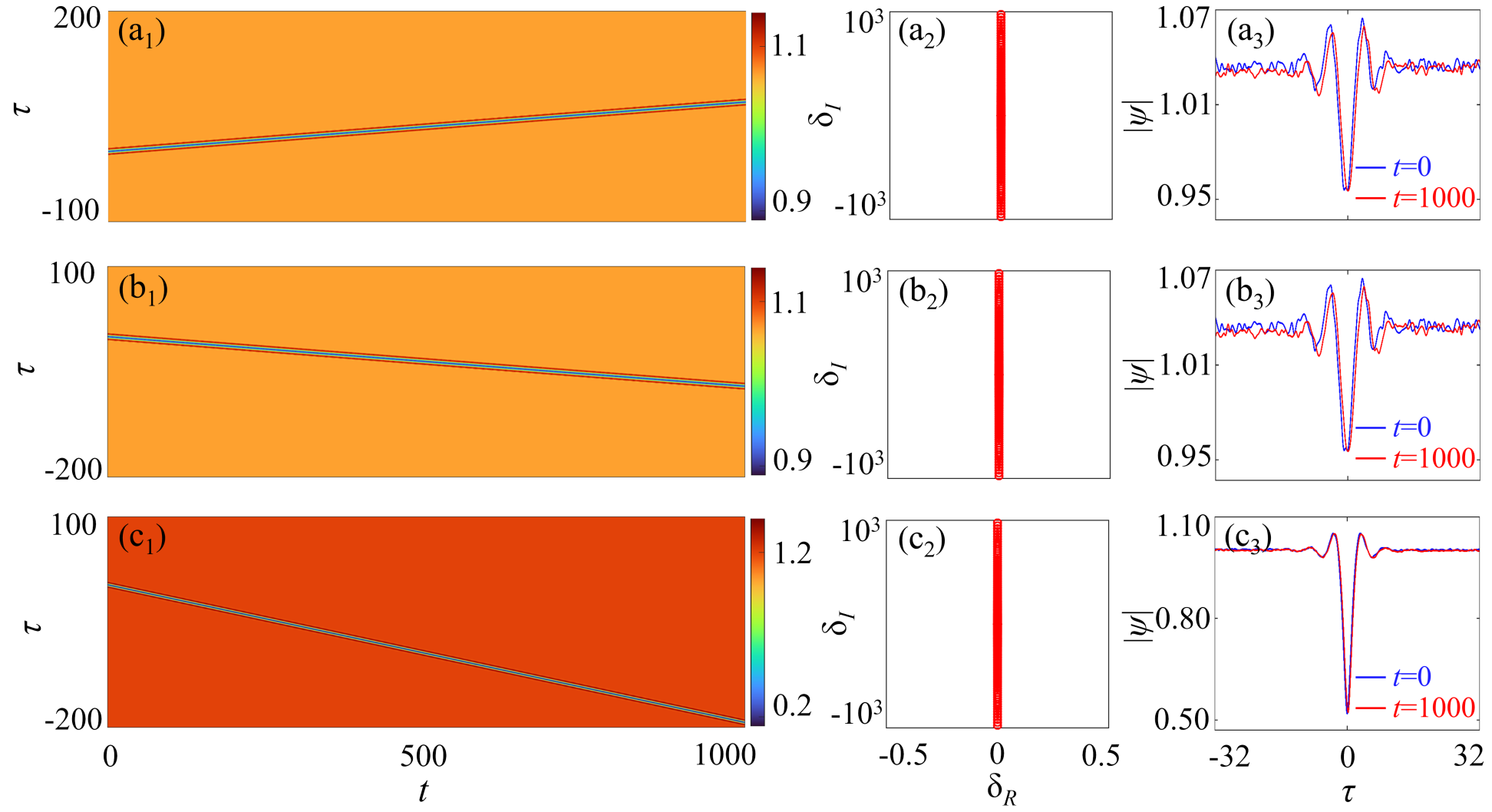} \vspace{%
			-0.5cm}
		\caption{The evolution and linear-stability spectra of MTSs
			for $m=4$ and $h=1/15$. The stable evolution of the MTSs subjected to $10\,\%$ random-amplitude noise for different soliton's velocities and momenta: $(\text{a}_{1})$ $V=0.05$ and $P=2.916$; $( \text{b}_{1})$ $V=-0.05$ and $P=-2.916$; $(\text{c}_{1})$ $V=-0.2$ and $P=0.776$. Panels $(\text{a}_{2})$, $(\text{b}_{2})$, and $(\text{c}_{2})$ give the associated stability spectra, while panels $(\text{a}_{3})$, $(\text{b}_{3})$, and $(\text{c}_{3})$ compare the perturbed input profiles with those obtained at $t=1000$.}
		\label{fig6}
	\end{figure}
	
	The typical dynamical behavior of the MTSs is displayed in Fig. \ref{fig6}.
	The trajectories in Figs. \ref{fig6}$(\text{a}_{1})$, \ref{fig6}$(%
	\text{b}_{1})$, and \ref{fig6}$(\text{c}_{1})$ show that the MTS profiles
	remain essentially unchanged up to $t=1000$. The comparison between Figs. %
	\ref{fig6}$(\text{a}_{1})$ and \ref{fig6}$(\text{b}_{1})$ demonstrates that
	two MTSs with opposite velocities undergo counterpropagating evolution. As
	illustrated in \ref{fig6}$(\text{c}_{1})$, the value of $|V|$ determines the
	drift speed of each soliton. The spectra in Figs. \ref{fig6}$(%
	\text{a}_{2})$, \ref{fig6}$(\text{b}_{2})$, and \ref{fig6}$(\text{c}_{2})$,
	obtained from Eq. (\ref{LSA-Eq}), are consistent with the stability for all
	three examples. Figures \ref{fig6}$(\text{a}_{3})$, \ref{fig6}$(\text{b}%
	_{3})$, and \ref{fig6} $(\text{c}_{3})$ compare the initial shapes of the
	perturbed soliton and final ones, produced by the extended evolution,
	further validating the robustness of the MTSs.
	
	\begin{figure}[h!]
		\centering\vspace{0cm} \includegraphics[width=1\linewidth]{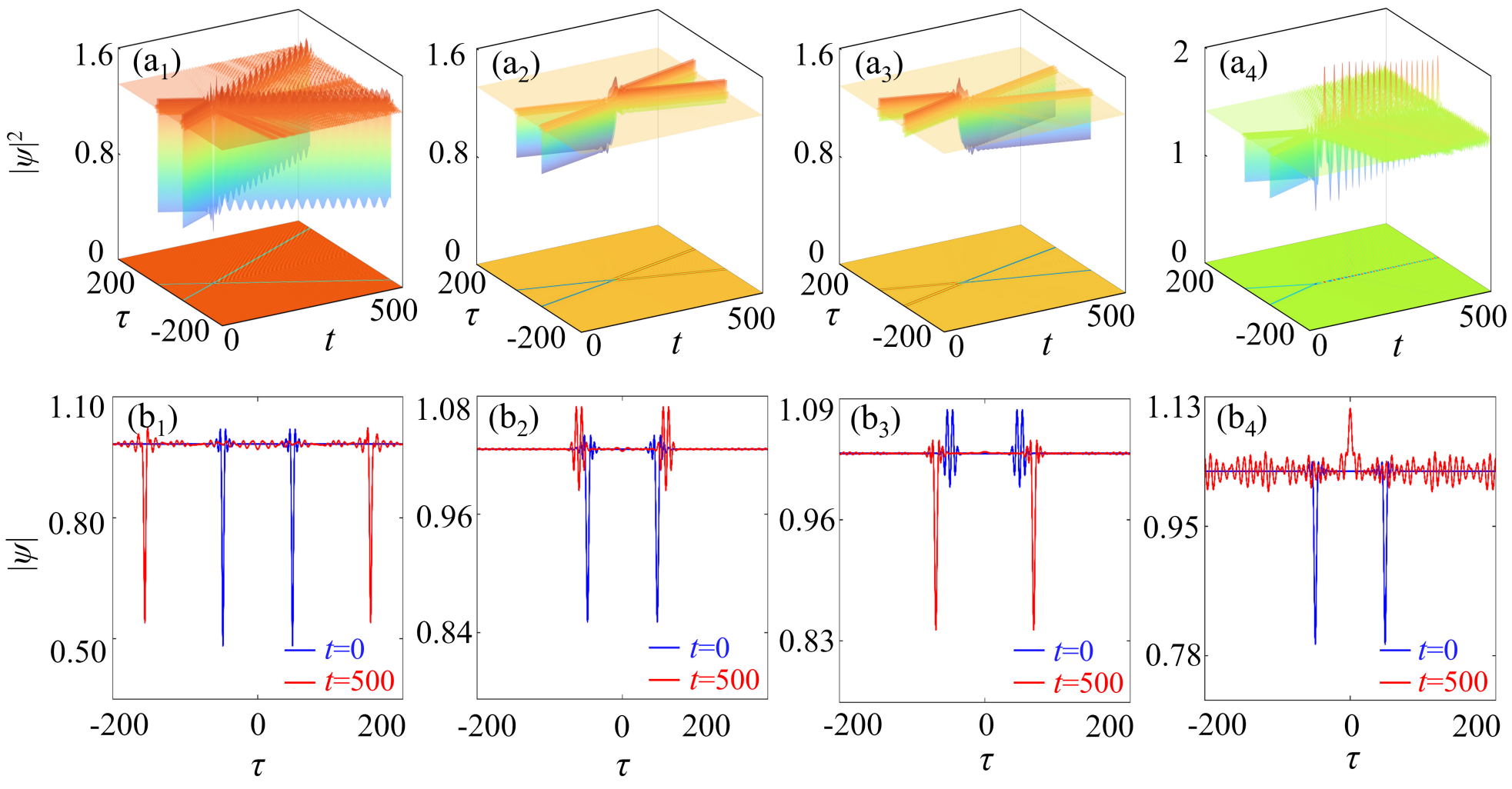} \vspace{%
			-0.5cm}
		\caption{Collisions between stable MTSs, with parameters $(
			\text{a}_{1})$ $V_{1}=-V_{2}=0.4$ and $P=1.717$; $(\text{a}_{2})$ $%
			V_{1}=-V_{2}=0.2$ and $P=2.680$; $(\text{a}_{3})$ $V_{1}=-V_{2}=0.24$ and $%
			P=3.310$; $(\text{a}_{4})$ $V_{1}=-V_{2}=0.3$ and $P=2.481$. Panels $(\text{%
				b }_{1})$, $(\text{b}_{2})$ , $(\text{b}_{3})$, and $(\text{b}_{4})$ display
			the initial MTS profiles and final ones, observed at $t=500$. Other
			parameters are $\protect\tau _{1}=\protect\tau _{2}=48$, $m=10$, and $h=1/15$.}
		\label{fig7}
	\end{figure}
	
	Finally, to explore interactions between MTSs, we construct the initial
	field profile for the head-on collision, as $\psi (\tau ,t=0)=\psi _{1}(\tau
	+\tau _{1})+\psi _{2}(\tau -\tau _{2})$, where the two solitons have
	opposite velocities, $V_{1}=-V_{2}$. Figure \ref{fig7}$(\text{a}_{1})$
	illustrates the repulsive soliton-soliton interaction, featuring rebound and
	emergence of pronounced amplitude oscillations, alongside emission of
	dispersive waves, as seen in Fig. \ref{fig7}$(\text{b}_{1})$. Figures \ref%
	{fig7}$(\text{a}_{2})$ and \ref{fig7}$(\text{a}_{3})$ present another
	repulsive regime, in which the collision triggers conspicuous deformation of
	the bouncing solitons, accompanied by the emission of weak dispersive-wave
	radiation. These phenomena reveal the inelastic character of the MTS
	collisions, as corroborated by the corresponding field profiles in Figs. \ref%
	{fig7}$(\text{b}_{2})$ and \ref{fig7}$(\text{b}_{3})$. Figure \ref{fig7}$(
	\text{a}_{4})$ depicts a different outcome of the collision, when the two
	MTSs undergo complete fusion, along with intense emission of dispersive waves,
	as seen in Fig. \ref{fig7}$(\text{b}_{4})$.
	
	\section{Conclusion}
	
	This work aims to produce dissipative dark-soliton families,
	sustained by the interplay of the PHEOD (pure-high-even-order dispersion) and parametric drive, extending the setting beyond the conventional second-order-dispersion regime. Two new families of stable topological solitons are found, namely, quiescent DTSs (dissipative topological solitons) and MTSs (moving topological solitons), for PHEOD orders $m=4,6,8$, and $10$. The topological solitons are combined with the zero-amplitude point, which connects two semi-infinite CW backgrounds with the $\pi $ phase shift between them. PHEOD gives rise to pronounced oscillatory tails, distinguishing these states from conventional dark solitons. The systematic linear-stability analysis and direct simulations corroborate the stability of both DTSs and MTSs. Pairwise collisions between the MTSs are inelastic, featuring two possibilities: the inter-soliton rebound or complete fusion, both accompanied by the emission of radiation waves.
	
	These results demonstrate that PHEOD can do more than smoothly distort conventional dissipative soliton profiles: it provides the mechanism for generating topological solitons with specific internal structures and interaction dynamics. These findings establish PHEOD as a basis for engineering robust dissipative topological solitons, both quiescent and moving ones. As an extension of the analysis, it may be interesting to construct nonlinear states in systems exhibiting competition between dispersion terms of different orders and different signs (normal/anomalous).
	
	\begin{backmatter}
		\bmsection{Funding}
		National Natural Science Foundation of China
		(Grant No. 62675244); Fundamental Research Program of Shanxi Province
		(Grant No. 202303021211185);  Israel Science Foundation (Grant No.
		1695/22); ANID (Chile) through FONDECYT (Grant No. 1260401).
		
		\bmsection{Disclosures}
		The authors declare no conflicts of interest.
		
		\bmsection{Data availability} Data underlying the results presented in this paper are not publicly available at this time but may be obtained from the authors upon reasonable request.
		
	\end{backmatter}
	
	
	
	

\end{document}